\documentstyle[aps,prl,twocolumn]{revtex}
\begin{document}
\draft
\title{Stability of a vortex in a trapped Bose-Einstein condensate}
\author{Anatoly A.~Svidzinsky and Alexander L.~Fetter}
\address{Department of Physics, Stanford University, Stanford, CA 94305-4060}
\date{\today }
\maketitle

\begin{abstract}
Based on the method of matched asymptotic expansion and on a time-dependent
variational analysis, we study the dynamics of a vortex in the
large-condensate (Thomas-Fermi) limit. Both methods as well as an analytical
solution of the Bogoliubov equations show that a vortex in a trapped
Bose-Einstein condensate has formally unstable normal mode(s) with positive
normalization and negative frequency, corresponding to a precession of the
vortex line around the center of the trap. In a rotating trap, the solution
becomes stable above an angular velocity $\Omega_m$ characterizing the onset
of metastability with respect to small transverse displacements of the
vortex from the central axis.
\end{abstract}

Recent experimental observations of Bose-Einstein condensation in trapped
alkali-atom gases at ultra-low temperatures \cite{And,Brad1,Dav} have raised
the question of vortex states similar to those in superfluid helium. Unlike
superfluid helium, however, the harmonic trap makes the alkali Bose
condensates nonuniform.

Numerical studies of a vortex in small to medium axisymmetric condensates
have determined the condensate density~\cite{Edw2,DS}, along with the
low-lying excitation spectrum~\cite{DBEC,IM}. These analyses found a
nontrivial eigenmode with positive norm and negative frequency, and a
variational treatment~\cite{DR1} suggested that such a system might be
unstable. In the Thomas-Fermi (TF) limit, the presence of a vortex \cite
{SS,SF,ZS} generally splits the eigenfrequencies of a vortex-free condensate
by a small amount of order $(d/R)^2\ll 1$ (but these analyses ignore the
possibility of the nonperturbative negative-frequency mode that occurs only
when a vortex is present). Separate studies have considered vortex formation
by a moving object~\cite{JCA} or a rotating force~\cite{MZ}.

The existence of a positive-norm eigenstate with negative-excitation energy~
\cite{DBEC,IM} formally implies an instability of the vortex state because
the condensate itself corresponds to zero energy. The present work obtains
explicit analytical negative-energy solutions for a vortex in a large
condensate, holding when the interatomic and trap interactions predominate
and the kinetic energy of the nonuniform density is small. This TF limit is
valid for $R/d\approx 15(Na/d)^{1/5}\gg 1$ \cite{BP}, where $R$ is the mean
condensate radius, $d=(d_xd_yd_z)^{1/3}$ is the mean harmonic oscillator
length [here, $d_j^2 =\hbar /M\omega _j$, with $M$ the atomic mass and $
\omega_j $ the corresponding trap frequency ($j =x, y, z$)], $N $ is the
number of atoms in the condensate, and $a>0$ is the $s$-wave scattering
length. Typically~\cite{Mewes} $d\approx $ a few $\mu m$, $a\approx $ a few
nm, and the experimental value $N\gtrsim 10^6$ amply satisfies the TF
criterion that $N\gg 10^3$. In the TF limit ($R\gg d$), the vortex core
radius $\xi \sim d^2/R$ is small compared to both $d$ and $R$.

In this paper we assume that most of the particles are in the condensate,
neglecting scattering with thermal background atoms, as confirmed by recent
experimental studies of the lowest-lying normal modes \cite{Jin,MOM} for low
temperatures. A condensate containing a $q$-fold quantized vortex in an
axisymmetric trap is known to have only stable normal density modes $\propto
e^{im\phi }$ for $m=0$ and $|m|\geq 2|q|$ (so that unstable modes occur only
for $0<|m|<2|q|$)~\cite{SF}. Here, we exhibit unstable solutions with $|m|=1$
that have positive norm and negative frequency in which the vortex line
precesses around the center of the trap.

{\it Dynamics of a vortex in the trapped condensate} ---Let us consider a
condensate in a nonaxisymmetric trap that rotates with an angular velocity $
\Omega $ around the $z$ axis. At zero temperature in a frame rotating with
the angular velocity $\Omega \hat z$, the evolution of the condensate wave
function $\Psi $ is described by the time-dependent Gross-Pitaevskii
equation \cite{LPP,EPG}:
\begin{equation}
\label{1}\left( -\frac{\hbar ^2}{2M}\nabla ^2+V_{{\rm tr}}+g|\Psi |^2-\mu
(\Omega )+i\hbar \Omega \partial _\phi \right) \Psi =i\hbar \frac{\partial
\Psi }{\partial t},
\end{equation}
where $V_{{\rm tr}}=\frac 12M\left( \omega _x^2x^2+\omega _y^2y^2+\omega
_z^2z^2\right) $ is the external trap potential, $g=4\pi \hbar ^2a/M>0$ is
the effective interparticle interaction strength, $\mu (\Omega )$ is the
chemical potential in the rotating frame, and $\phi $ is the azimuthal angle
in cylindrical polar coordinates. We assume that the condensate contains a $
q $-fold quantized vortex parallel to the $z$ axis located near the center
of the trap at the transverse position ${\bbox \rho }_0(t)$. In this section
we use the method of matched asymptotic expansion to determine the vortex
velocity as a function of the local gradient of the trap potential and
angular velocity $\Omega $, generalizing two-dimensional results obtained by
Rubinstein and~Pismen \cite{RP} to the case of a three-dimensional rotating
harmonic potential. The method applies when the external potential does not
change significantly on distances comparable with the core size $|q|\xi \ll
R_{\perp }$ (TF\ limit); it matches the outer asymptotics of the solution of
Eq.~(\ref{1}) in the vortex core region ($|{\bbox \rho }-{\bbox
\rho }_0|\leq |q|\xi $) with the short-distance behavior of the solution in
the region far from the vortex core ($|{\bbox \rho }-{\bbox \rho }_0|\gg
|q|\xi $). Recently the method was used to analyze the drift of an optical
vortex soliton in a slowly varying background field \cite{KCDTP}.

To find the solution in the vortex core region, one may consider Eq.~(\ref{1})
in a coordinate frame centered on the vortex line that moves with the
vortex velocity ${\bf V}$ (${\bf V}\perp \hat z$). The solution is assumed
to be stationary in the comoving frame and satisfies the equation:
$$
\left( -\frac{\hbar ^2}{2M}\nabla ^2+V_{{\rm tr}}+g|\Psi |^2-\mu (\Omega
)+i\hbar \Omega \partial _\phi \right) \Psi =
$$
\begin{equation}
\label{2}=-i\hbar {\bf V}\cdot {\bf \nabla }_{\perp }\Psi \enspace .
\end{equation}
In the vortex core region we may seek a solution in the form of an expansion
in the small parameter $\xi /R_{\perp }$:
\begin{equation}
\label{3}\Psi =\left[ |\Psi _0(\rho ,z)|-\chi (\rho ,z)\cos \phi \right]
e^{iq\phi -i\eta (\rho ,z)\sin \phi },
\end{equation}
where $\Psi _0$ is the condensate wave function with $V_{{\rm tr}}$ replaced
by $V_z=\frac 12M\omega _z^2z^2$ and $\chi $ and $\eta $ characterize the
perturbation in the absolute value and phase. Physically $\Psi _0$ is the
analogous wave function for an unbounded condensate in the $xy$ plane with
the same chemical potential $\mu (\Omega )$. The polar angle $\phi $ is
measured from the direction of the local potential gradient ${\bf \nabla }
_{\perp }V_{{\rm tr}}({\bbox \rho }_0)$, and $\rho $ is the local radial
cylindrical coordinate. For $\rho \gg |q|\xi $ the perturbations have the
asymptotic form:

\begin{equation}
\label{4}\eta \approx \frac{q|{\bf \nabla }_{\perp }V_{{\rm tr}}({\bbox \rho
}_0)|}{2g|\Psi _0|^2}\rho \ln \left( A\rho \right) ,\enspace
\chi \approx \frac{|{\bf \nabla }_{\perp }V_{{\rm tr}}({\bbox \rho }_0)|}{
2g|\Psi _0|}\rho .
\end{equation}
Here we omit terms containing $\Omega $ explicitly because they are
proportional to the small parameter $\hbar \Omega /\mu $. The parameter $A$
must be determined by matching the solutions (\ref{4}) with those far from
the vortex core; in fact, $A$ depends on $\Omega $.

To lowest order in the small parameter $\xi /R_{\perp }$, Eq.~(\ref{1}) far
from the vortex core reduces to an equation for the condensate phase only ($
\Psi =|\Psi |e^{iS}$):
\begin{equation}
\label{5}|\Psi _{TF}|^2\nabla ^2S+\nabla |\Psi _{TF}|^2\cdot\nabla S-\frac{
M\Omega }\hbar \partial _\phi |\Psi _{TF}|^2=0,
\end{equation}
where $g|\Psi _{TF}|^2\approx \mu -\frac 12M\omega _z^2z^2$ and $\mu \equiv
\mu (\Omega )+$ $\hbar \Omega q$. Comparing the solution of Eq. (\ref{5})
for $\rho \ll R_{\perp }$ with the outer asymptotic form (\ref{4}) of the
core solution gives the value of the parameter $A$. With logarithmic
accuracy we find
\begin{equation}
\label{6}\ln \left( Ae\right) =\ln \left( \frac 1{R_{\perp }}\right) +\frac{
4\Omega g|\Psi _0|^2}{q\hbar (\omega _x^2+\omega _y^2)},
\end{equation}
where $R_{\perp }$ is the mean transverse dimension of the condensate. To
find the velocity ${\bf V}$ of the vortex, one can use a solvability
condition (Fredholm's alternative) of the first-order equation in the small
parameter (see \cite{RP} for details). In the frame rotating with the trap,
the vortex velocity is found to be orthogonal to the direction of the local
gradient of the trap potential ${\bf \nabla }_{\perp }V_{{\rm tr}}$:
\begin{equation}
\label{7}{\bf V=}\frac{3q\hbar }{4M\mu }\left[ \ln \left( \frac{R_{\perp }}{
|q|\xi }\right) -\frac{8\mu \Omega }{3q\hbar (\omega _x^2+\omega _y^2)}
\right] \left( \hat z\times {\bf \nabla }_{\perp }V_{{\rm tr}}\right) ,
\end{equation}
where $\mu $ can be taken as the chemical potential for a nonrotating trap.
The vortex follows an elliptic trajectory along the line $V_{{\rm tr}}=const$.
For $q>0$ and $\Omega <\Omega _m=3|q|\hbar \left( \omega _x^2+\omega
_y^2\right) \ln \left( R_{\perp }/|q|\xi \right) /8\mu $, the vortex moves
counter-clockwise in the positive sense. With increasing rotation frequency $
\Omega $ of the trap, the vortex velocity (as seen in the rotation frame)
decreases towards zero and vanishes at $\Omega =\Omega _m$. At this rotation
frequency, the effective potential for the vortex becomes flat near the trap
center [see Eq.~(\ref{18}) below]. For $\Omega >\Omega _m$ the apparent
motion of the vortex becomes clockwise.

From Eq.~(\ref{7}) it is straightforward to obtain the vortex position as a
function of time:
\begin{equation}
\label{8}x_0=\varepsilon _0R_x\sin \left( \omega t+\phi _0\right) ,
\end{equation}
\begin{equation}
\label{9}y_0=\pm \varepsilon _0\,R_y\cos \left( \omega t+\phi _0\right) ,
\end{equation}
where
\begin{equation}
\label{10}\omega =\pm \left[ \frac{2\omega _x\omega _y\Omega }{(\omega
_x^2+\omega _y^2)}-\frac{3q\hbar \omega _x\omega _y}{4\mu }\ln \left( \!
\frac{R_{\perp }}{|q|\xi }\right) \right] \,.
\end{equation}
This vortex motion near the trap center can be considered a collective
excitation of the condensate with an eigenfrequency $\omega $. The two
different frequencies given by (\ref{10}) correspond to solutions with
different sign of normalization: in terms of the Bogoliubov amplitudes $u(
{\bf r}),v({\bf r})$, these solutions have the normalization $\int
d^3r\left( |u^{\pm }|^2-|v^{\pm }|^2\right) =\pm {\rm sgn}(q)$. The solution
with positive norm has the energy
\begin{equation}
\label{12}E={\rm sgn}(q)\frac{2\omega _x\omega _y\hbar \Omega }{(\omega
_x^2+\omega _y^2)}-\frac{3|q|\hbar ^2\omega _x\omega _y}{4\mu }\ln \left(
R_{\perp }/|q|\xi \right) .
\end{equation}
If $\Omega =0$, the excitation energy is negative and hence formally
unstable. This negative energy implies the existence of a state with lower
energy that can become macroscopically occupied. Nevertheless, the
condensate will be unstable only if there is a mechanism to transfer the
system to the lower energy state \cite{Pu}, which requires the presence of a
reservoir to conserve energy and angular momentum. Thus a vortex in a
nonrotating harmonic trap can remain stable at low enough temperatures when
the noncondensate atoms are negligible. If the trap rotates, Eq.~(\ref{12})
shows that the solution becomes stable at $|\Omega |\geq $ $\Omega _m$ which
coincides with the angular velocity at which the vortex at the center
becomes metastable (see below). Note that the precession frequency (10) in a
nonrotating trap involves the product $\omega _x\omega _y$, whereas the
metastable rotation frequency $\Omega _m$ instead involves the combination $
\frac 12(\omega _x^2+\omega _y^2)$.

{\it Analysis of the Bogoliubov equations for an axisymmetric trap} ---  
The collective excitation
energies $E$ of the (in general, nonuniform) condensate are the eigenvalues
of the Bogoliubov equations for the coupled amplitudes $u({\bf r})$, $v({\bf
r})$
$$
\left( -\frac{\hbar ^2}{2M}\nabla ^2+V_{{\rm tr}}+2g|\Psi |^2-\mu (\Omega
)\right) \left(
\begin{array}{c}
u \\
v
\end{array}
\right)
$$
\begin{equation}
\label{b2}+\left(
\begin{array}{cc}
i\hbar \Omega \partial _\phi  & -g\Psi {}^2 \\
-g\Psi ^{*}{}^2 & -i\hbar \Omega \partial _\phi
\end{array}
\right) \left(
\begin{array}{c}
u \\
v
\end{array}
\right) =E\left(
\begin{array}{c}
u \\
-v
\end{array}
\right) .
\end{equation}

In this section we assume the trap potential to be axisymmetric, that is $
\omega _x=\omega _y=\omega _{\perp }$. We can rewrite the Bogoliubov
equations in the following form:
\begin{equation}
\label{b3}\hat H_0\left(
\begin{array}{c}
u \\
v
\end{array}
\right) +\hat V\left(
\begin{array}{c}
u \\
v
\end{array}
\right) =E\left(
\begin{array}{c}
u \\
-v
\end{array}
\right) ,
\end{equation}
where
$$
\hat H_0=\left( -\frac{\hbar ^2}{2M}\nabla ^2+\frac 12M\omega
_z^2z^2+2g|\Psi _0|^2-\mu (\Omega )\right) \!\left(
\begin{array}{cc}
1 & 0 \\
0 & 1
\end{array}
\right)
$$
\begin{equation}
\label{b4}+\left(
\begin{array}{cc}
i\hbar \Omega \partial _\phi  & -g\Psi _0{}^2 \\
-g\Psi _0^{*}{}^2 & -i\hbar \Omega \partial _\phi
\end{array}
\right) ,
\end{equation}
$$
\hat V=\left( \frac 12M\omega _{\perp }^2\left( x^2+y^2\right) +2g\left(
|\Psi |^2-|\Psi _0|^2\right) \right) \left(
\begin{array}{cc}
1 & 0 \\
0 & 1
\end{array}
\right)
$$
\begin{equation}
\label{b5}+\left(
\begin{array}{cc}
0 & g\left( \Psi _0{}^2-\Psi ^2\right)  \\
g\left( \Psi _0^{*}{}^2-\Psi ^{*2}\right)  & 0
\end{array}
\right) ,
\end{equation}
and $\Psi _0$ is the wave function for an unbounded condensate in the $xy$
plane with the same chemical potential $\mu (\Omega )$; its excitations obey
the equation
\begin{equation}
\label{b6}\hat H_0\left(
\begin{array}{c}
u_0 \\
v_0
\end{array}
\right) =E_0\left(
\begin{array}{c}
u_0 \\
-v_0
\end{array}
\right) .
\end{equation}

For a $q$-fold quantized vortex at the center of the trap parallel to the $z$
-axis, the unperturbed condensate wave function has the form $\Psi
_0=e^{iq\phi }|\Psi _0|$, and Eq.~(\ref{b6}) has the following exact pair of
solutions~\cite{SF}
\begin{equation}
\label{b7}U_{\pm }=\left(
\begin{array}{c}
u_0^{\pm } \\
v_0^{\pm }
\end{array}
\right) =\frac{e^{\pm i\phi }}I\left(
\begin{array}{c}
{e^{iq\phi }}\left( -\partial _\rho \pm q/\rho \right) \\ {e^{-iq\phi }}
\left( \partial _\rho \pm q/\rho \right) ,
\end{array}
\right) |\Psi _0|,
\end{equation}
\begin{equation}
\label{b8}E_0=\mp \hbar \Omega ,
\end{equation}
where $\rho $ is the radial cylindrical coordinate, and $I^2=4\pi |q|\int
dz\left| \Psi _0(\rho =\infty )\right| ^2\approx 16\sqrt{2}\pi |q|\mu
^{3/2}/3g\omega _z\sqrt{M}$ in the TF limit. These solutions are normalized
so that $\int d^3r\left( |u_0^{\pm }|^2-|v_0^{\pm }|^2\right) ={\rm sgn}(\mp
q)$. They generalize those obtained by Pitaevskii~\cite{LPP} to the case of
an axial trap with potential $V_z$ and describe the transverse motion of the
vortex line as a whole relative to the condensate. It is obvious that the
solutions $U_{+}$ and $U_{-}$ are orthogonal.

We use Eqs.~(\ref{b7})-(\ref{b8}) as a zero-order approximation for a
perturbation expansion of Eq.~(\ref{b3}). The first-order shift of $E$
requires the matrix elements $\langle U_{\pm }|\hat V|U_{\pm }\rangle $.
With logarithmic accuracy, their main contribution arises from the
integration region $|q|\xi \ll \rho \ll R_{\perp }$, where $|q|\xi $ is the
vortex core radius. For this region in the TF limit, we have
\begin{equation}
\label{b9}g|\Psi |^2\approx \mu -\frac{\hbar ^2q^2}{2M\rho ^2}-\!V_{{\rm tr}
}\!-\frac{\hbar ^2\!\omega _{\perp }^2}{2g|\Psi _{TF}|^2}+\frac{\hbar ^2}{2M}
\frac{\partial _{zz}^2|\Psi _{TF}|}{|\Psi _{TF}|},
\end{equation}
where $g|\Psi _{TF}|^2\approx \mu -\frac 12M\omega _z^2z^2$ and $\mu \equiv
\mu (\Omega )+$ $\hbar \Omega q$. The last two terms in (\ref{b9}) reflect
the gradient terms in the Gross-Pitaevskii equation and are small in the TF
limit. Nevertheless, these small terms give the main contribution to the
matrix elements. The same expression (\ref{b9}) holds for the density of the
unbounded condensate $|\Psi _0|^2$ but with $\omega _{\perp }=0$ . Thus the
perturbation operator $\hat V$ becomes
$$
\hat V\!=\!\left( \!\frac 12M\omega _{\perp }^2\!\left( x^2\!+\!y^2\right)
\!+\!\frac{\hbar ^2\omega _{\perp }^2}{2g|\Psi _{TF}|^2}\!\right) \!\!\left(
\!
\begin{array}{cc}
-1 & {e^{2iq\phi }} \\ {e^{-2iq\phi }} & -1
\end{array}
\!\right)
$$
\begin{equation}
\label{b10}-\frac{\hbar ^2\omega _{\perp }^2}{2g|\Psi _{TF}|^2}\left(
\begin{array}{cc}
1 & 0 \\
0 & 1
\end{array}
\right) ,
\end{equation}
and the off-diagonal matrix elements $\left\langle U_{\pm }|\hat V|U_{\mp
}\right\rangle $ vanish. With logarithmic accuracy, the last term in
Eq.~(\ref{b10}) gives the main contribution to
\begin{equation}
\label{b11}\left\langle U_{\pm }|\hat V|U_{\pm }\right\rangle \approx -\frac{
\hbar ^2\omega _{\perp }^2}{gI^2}\int d^3r\frac{q^2}{\rho ^2}.
\end{equation}
Integrating over $d\rho $ gives a logarithmic factor $\ln \left( R_{\perp
}/|q|\xi \right) $.

As a result, we obtain two solutions for the excitation energy. They are
normalized to the value ${\rm sgn}(\mp q)$, and the solution with positive
norm has an energy
\begin{equation}
\label{b12}E={\rm sgn}(q)\hbar \Omega -\frac{3|q|\hbar ^2\omega _{\perp }^2}{
4\mu }\ln \left( R_{\perp }/|q|\xi \right) ,
\end{equation}
where $\mu $ can be taken as the chemical potential for a nonrotating trap.
Formula (\ref{b12}) agrees with (\ref{12}) obtained on the basis of the
method of matched asymptotic expansion (for $\omega _x=\omega _y=\omega
_{\perp }$).

{\it Energy of a vortex in a Bose-Einstein condensate in a rotating trap }
--- In a frame rotating with angular velocity $\Omega \hat z$, the energy
functional of the system is
\begin{equation}
\label{13}E(\Psi )\!=\!\!\int\!\! dV\!\left(\! \frac{\hbar ^2}{2M}|\nabla
\Psi |^2\!+V_{{\rm tr}}|\Psi |^2\!+\frac{g}{2}|\Psi |^4\!+\Psi ^{*}i\hbar
\Omega \partial _\phi \Psi \!\right)\! .
\end{equation}
Consider a $q$-fold quantized vortex displaced from the center of the trap,
with transverse coordinates $x=x_0$, $y=y_0$. In the TF limit, the
vortex-induced change in the condensate density is negligible. Hence, the
main contribution to the change $\Delta E(x_0,y_0)$ in the energy of the
system due to the vortex arises from the condensate's superfluid motion, and
we can take $|\Psi |^2$ in Eq.~(\ref{13}) to be the density of the
condensate without a vortex. A transverse shift of the Cartesian coordinate
system places the origin on the vortex axis and yields
\begin{equation}
\label{14}\Psi =|\Psi |e^{iq\phi +iS_0}\enspace,
\end{equation}
\begin{equation}
\label{15}g|\Psi |^2=\mu -\frac 12M\left[ \omega _x^2(x+x_0)^2+\omega
_y^2(y+y_0)^2+\omega _z^2z^2\right] \enspace,
\end{equation}
where $\phi $ is a polar angle around the vortex line and $S_0$ is a
periodic function of $\phi $.

Varying the functional (\ref{13}) gives an Euler-Lagrange equation for $S_0$
. For a vortex-free condensate, we get
\begin{equation}
\label{16}S_0=-\frac{M\Omega }\hbar \left(\frac{ \omega _x^2-\omega _y^2 }{
\omega _x^2+\omega _y^2 }\right)(x+x_0)(y+y_0)\enspace.
\end{equation}
If the condensate has a $q$-fold quantized vortex, then the contribution to
the free energy from the phase factor $iS_0$ in Eq.~(\ref{14}) can be
neglected with respect to the contribution from $iq\phi $ for small angular
velocity $\Omega $. Only for $|\Omega |\gg |q|\hbar \left( \omega
_x^2+\omega _y^2\right) /4\mu $ do the contributions from these two phase
factors become comparable. For such large angular velocities, however, the
function $S_0$ is insensitive to the presence of the vortex. Thus Eq.~(\ref
{14}) with $S_0$ given by Eq.~(\ref{16}) provides a good approximation for
the condensate wave function for general angular velocity $\Omega$.

Substituting Eqs.~(\ref{14})-(\ref{16}) into the energy functional (\ref{13})
 and integrating with logarithmic accuracy over the volume of condensate
yield the following expression
$$
\Delta E(x_0,y_0)=\frac{8\pi }3\mu R_z\xi ^2n_0(0)\left( 1-\frac{x_0^2}{
R_x^2 }-\frac{y_0^2}{R_y^2}\right) ^{3/2}
$$
\begin{equation}
\label{17}\times \bigg[ q^2\ln \left( \frac{R_{\perp }}{|q|\xi }\right) -
\frac{8q\mu \Omega }{5\hbar \left( \omega _x^2+\omega _y^2\right) }\left( 1-
\frac{x_0^2}{R_x^2}-\frac{y_0^2}{R_y^2}\right) \bigg],
\end{equation}
where $R_j^2=2\mu /M\omega _j^2$ fixes the condensate's dimensions, $
R_{\perp }^{-2}=M(\omega _x^2+\omega _y^2)/4\mu =\frac{1}{2}
(R_x^{-2}+R_y^{-2})$ 
characterizes the mean transverse dimension, $\xi ^2=\hbar ^2/2M\mu $, and $
n_0(0)=\mu /g$ is the density at the center of the vortex-free condensate.
For small displacements of the vortex ($x_0\ll R_x$, $y_0\ll R_y$), we have:
$$
\Delta E(x_0,y_0)\!=\frac{8\pi }3\mu R_z\xi ^2n_0(0)\!\bigg\{ q^2\ln
\!\left( \frac{R_{\perp }}{|q|\xi }\right) -\frac{8q\mu \Omega }{5\hbar
\left( \omega _x^2\!+\!\omega _y^2\right) }
$$
\begin{equation}
\label{18}-\frac 12\left( \frac{x_0^2}{R_x^2}+\frac{y_0^2}{R_y^2}\right)
\bigg[ 3q^2\ln \left( \frac{R_{\perp }}{|q|\xi }\right) -\frac{8q\mu \Omega
}{\hbar \left( \omega _x^2+\omega _y^2\right) }\bigg]\bigg\}.
\end{equation}
For $|\Omega |>\Omega _{c}=5|q|\hbar \left( \omega _x^2+\omega _y^2\right)
\ln \left( R_{\perp }/|q|\xi \right) /8\mu =\frac 52|q|\hbar /(MR_{\perp
}^2)\ln \left( R_{\perp }/|q|\xi \right) $, the central vortex is stable
because $\Delta E(0,0)$ is negative~\cite{LPS}. Furthermore, $\Delta
E(x_0,y_0)$ has a local minimum at the origin for $|\Omega |>\Omega _m=\frac
35\Omega _{c}$, showing that the central vortex is metastable in the
interval $\Omega _m<|\Omega |<\Omega _{c}$. Since $\hbar \omega _{x,y,z}\ll
\mu $ in the TF limit, the thermodynamic critical angular velocity $\Omega
_{c}$ for vortex formation and $\Omega _m$ are both much smaller than the
trap frequencies.

{\it Time-dependent variational analysis} --- Instead of solving the
time-dependent Gross-Pitaevskii equation~(\ref{1}) directly, one may
consider a variational problem for the action obtained from the effective
Lagrangian (we assume $\Omega =0$ in this section):
$$
L(t)=\int d^3r\bigg[ \frac i2\hbar \left( \Psi ^{*}\frac{\partial \Psi }{
\partial t}-\Psi \frac{\partial \Psi ^{*}}{\partial t}\right)
$$
\begin{equation}
\label{20}-\frac{\hbar ^2}{2M}\left| \nabla \Psi \right| ^2-V_{{\rm tr}}(
{\bf r})\left| \Psi \right| ^2-\case{1}{2}g\left| \Psi \right| ^4\bigg] .
\end{equation}
With a suitable trial function containing time-dependent parameters, the
time evolution follows from the associated Lagrange equations for these
parameters. Although not exact, this method gives a clear physical picture
of the system's behavior. It determined the low-energy excitations of a
trapped zero-temperature condensate~\cite{PMCLZ}, predicting the low-energy
excitation spectrum of the condensate and, in the TF limit, reproducing the
formulas derived by Stringari~\cite{Str}. In the present paper we use this
method as an alternative way to study the normal modes of a condensate with
a vortex.

We are interested in the relative motion of a vortex inside the condensate,
so it is natural to take a time-dependent trial function in the form:
$$
\Psi \left( {\bf r},t\right) =B(t)f\left[ {\bf r}-{\bbox \rho}_0(t)\right]
F[ {\bf r}-{\bbox \eta }_0(t)]
$$
\begin{equation}
\label{21}\times \prod_{j =x,y,z}\exp \left[ ix_j \alpha _j (t)+ix_j ^2\beta
_j (t)\right] .
\end{equation}
Here the function $f\left({\bf r}\right) $ characterizes the vortex inside
the trap and has the approximate form $f\left( {\bf r}\right) =e^{iq\phi }$
for $\rho \gg |q|\xi$; the function $F({\bf r})$ describes the condensate
density distribution and, in the TF limit, has the form
\begin{equation}
\label{22}F^2({\bf r})=1-\frac{x^2}{R_x^2(t)}-\frac{y^2}{R_y^2(t)} -\frac{
z^2 }{R_z^2(t)}.
\end{equation}
The time-dependent vector ${\bbox \eta }_0(t)=(\eta _{0x},\eta _{0y},\eta
_{0z})$ describes the motion of the center of the condensate, while ${\bbox
\rho }_0(t)=(x_0,y_0,0)$ describes the motion of the vortex line in the $xy$
plane. The other variational parameters are the amplitude $B$, the width of
the condensate $R_j $, and the set $\alpha _j $, and $\beta _j $ that
describe the condensate's motion. These parameters are real functions of
time, characterizing the macroscopic wave function of the condensate.
Substitution of the trial wave function (\ref{21}) into Eq.~(\ref{20})
yields an effective Lagrangian as a function of the variational parameters
(and their first time derivatives).

We focus on small oscillations, when the displacement ${\bbox \rho }_0(t)$
of the vortex from the trap center and the displacement ${\bbox \eta }_0(t)$
of the condensate are small relative to the width of the condensate;
expanding for small ${\bbox \rho }_0(t)$ and ${\bbox \eta }_0(t)$, we keep
only the first nonzero corrections. The resulting equations for the motion
of the vortex line (that is, for $x_0$, $y_0$) are coupled only with $\eta
_{0x},\eta _{0y}$ and $\alpha _x,\alpha _y$. This system of six first-order
ordinary differential equations can be solved explicitly. Apart from the
well-known dipole-mode solutions that describe the rigid oscillation of the
condensate and the vortex as a whole, these equations have another solution
that corresponds to the motion of the vortex relative to the condensate. For
this solution the vortex motion is described by formulas (\ref{8}), (\ref{9}),
 while the displacement of the condensate is given by
\begin{equation}
\label{25}\eta _{0x}=-\frac{15\varepsilon _0\xi ^2}{2R_y}\ln \left( \!\frac{
R_{\perp }}{|q|\xi }\right) \frac{R_x}{R_x+R_y}\sin \left( \omega t+\phi
_0\right) ,
\end{equation}
\begin{equation}
\label{26}\eta _{0y}=\mp \frac{15\varepsilon _0\xi ^2\,}{2R_x}\ln \left( \!
\frac{R_{\perp }}{|q|\xi }\right) \frac{R_y}{R_x+R_y}\cos \left( \omega
t+\phi _0\right) ,
\end{equation}
where
\begin{equation}
\label{27}\omega =\mp \frac{3q\hbar \omega _x\omega _y}{4\mu }\ln \left( \!
\frac{R_{\perp }}{|q|\xi }\right) \,.
\end{equation}

The quantity $x_0^2/R_x^2+y_0^2/R_y^2=\varepsilon _0^2$ remains constant as
the vortex line follows an elliptic trajectory around the center of a trap
along the line $V_{{\rm tr}}=const$, and the energy of the system is
conserved [as follows from Eq.~(\ref{17})]. These results agree with those
obtained from the method of matched asymptotic expansion. The condensate
also precesses with relative phase shift $\pi $ at the same frequency, but
the amplitude of the condensate motion is smaller than that of the vortex
line by a factor $\sim \xi ^2\ln \left( R_{\perp }/|q|\xi \right) /R_xR_y$.
As in the discussion of (10), these solutions have the normalization $-{\rm
sgn}(\omega )$, and the positive normalization corresponds to the solution
with negative frequency. In the present limit of logarithmic accuracy, the
structure of the vortex core is inessential for the excitation spectrum. The
precession frequency (\ref{27}) obtained in the variational approach
coincides with formula (\ref{10}) (for $\Omega =0$).

In conclusion, we have used the method of matched asymptotic expansion (for
a general three-dimensional rotating nonaxisymmetric trap), analysis of the
Bogoliubov equations and a time-dependent variational procedure to study
particular normal modes of a trapped condensate containing a vortex. These
modes have negative energy and positive norm, with the vortex line moving
around the $z$ axis. The excitation frequency in a rotating trap becomes
positive when the angular velocity $\Omega $ reaches the value that makes
the central vortex metastable.

This work was supported in part by the National Science Foundation, Grant
Nos. DMR 94-21888 and 99-71518, and by Stanford University (A.A.S.).

\end{document}